\documentclass[aps,twocolumn,prl,superscriptaddress,showpacs,preprintnumbers,amsmath,amssymb]{revtex4}

\usepackage{graphicx}
\usepackage{dcolumn}
\usepackage{bm}

\begin{document}

\preprint{APS/123-QED}


\title{Multiple-time scaling and universal behaviour of the earthquake inter-event
  time distribution}

\author{M. Bottiglieri}
 \affiliation{Department of Environmental Sciences and CNISM, Second University of
Naples, Caserta, Italy.}

\author{L. de Arcangelis}
\affiliation{IfB, ETH, Schafmattstr.6, 8093 Z\"urich, CH;
Department of Information Engineering and CNISM, Second University of
Naples, Aversa (CE), Italy}%

\author{C. Godano}
\affiliation{Department of Environmental Sciences and CNISM, Second
  University of 
Naples, Caserta, Italy.}

\author{E. Lippiello}
 \affiliation{Department of Environmental Sciences and CNISM, Second University of
Naples, Caserta, Italy.}

\date{\today}

\begin{abstract}
The inter-event time distribution 
characterizes the
temporal occurrence in seismic catalogs. Universal scaling
properties of this distribution have been evidenced for entire catalogs and
seismic sequences. Recently,
these universal features
have been questioned and
some criticisms have been raised. We investigate the existence of
universal  scaling properties
by analysing a Californian catalog and by means of numerical
simulations of an epidemic-type model. 
We show that the inter-event time distribution exhibits
a universal behaviour over the entire temporal range
 if four characteristic times are taken into account.
The above analysis allows to identify the scaling form leading to
universal behaviour and explains the observed deviations.
Furthermore it provides a tool 
to identify the dependence  on mainshock magnitude of the $c$
parameter that  fixes the onset of
the power law decay in the Omori law.

\end{abstract}

\maketitle


Seismic occurrence is a phenomenon of great complexity involving
different processes acting on different time and space scales. 
In the last decade a unifying picture of seismic occurrence has been
proposed via the investigation of $D(\Delta t)$, the distribution of
inter-event times $\Delta t$ between successive earthquakes
\cite{bak,cor,cor1}.  
These studies   
have shown that, rescaling inter-event times by the average occurrence
rate 
$R$, $D(\Delta t)$  follows the scaling relation
\begin{equation}
D(\Delta t)=R f(R \Delta t)
\label{corral}
\end{equation}
where the functional form of $f(x)$ is quite independent of the
geographic  zone and the magnitude threshold. 
The above relation
suggests that $R$ is a non universal quantity and is the only
typical inverse time scale affecting  $D(\Delta t)$.
This result, obtained for periods of stationary rate,
has been generalized to
non stationary periods \cite{cor2} and Omori sequences
\cite{shc,bot}.
The scaling relation (\ref{corral}) has been also observed  for volcanic 
earthquakes
\cite{bot1}.
On the other hand, recent studies  have questioned the universality of
the inter-event time distribution
\cite{dav,mol,hai,ss,ss2,ss3,tou}. 
In particular, deviations from universality at small  $\Delta t$ 
have been related to the interplay between
correlated earthquakes, following a gamma distribution, and
uncorrelated events, following a pure exponential decay \cite{tou}.
 This behaviour is well described by  numerical
simulations of the Epidemic Type Aftershock Sequence (ETAS) model \cite{oga}.
 Indeed, analytical studies  \cite{mol} and
a previous numerical  analysis of the ETAS model \cite{hai},
have shown that the functional form of  $D(\Delta t)$
depends on the ratio between correlated and independent
earthquakes ${\cal K}$. The problem has been also attacked
within the theoretical framework of probability generating   
functions \cite{ss,ss2,ss3}. Saichev \& Sornette (SS)  
have  obtained an exact nonlinear integral equation for the
ETAS model and solved it analytically at linear order \cite{ss,ss2}. 
This solution
shows  that the function $f(x)$ in Eq.(\ref{corral}) depends on ${\cal
K}$ and on some other parameters of the model. This behavior is confirmed if
non-linear  contributions are taken into account \cite{ss3}.       

Multiplicity of characteristic times is
often observed in the dynamics of complex systems, where different
temporal scales are associated to the relaxation of different spatial
regions or structures. For instance, their existence is a well
established property in glassy materials, polymers or gelling
systems, where they originate from the relaxation of complex
structures at different mesoscopic scales, or else from the
emergence of competing interactions \cite{ang}.
Moreover, the coexistence of different
physical mechanisms acting at different spatio-temporal scales 
may also give rise
to complex temporal scaling \cite{tar}.
Therefore, the identification of the number of relevant 
time scales controlling universal behaviours 
is a very debated subject in complex systems.

In this paper, we do not assume
the existence of a unique time scale $1/R$, as in Eq.(\ref{corral}). 
We show that four typical timescales are relevant for the
inter-event time distribution scaling:  the inverse rate of  independent
events $\mu$, the average inverse rate of correlated events,
the time parameter $c$ defined in the
Omori law and the catalog duration $T$.
 These different time scales lead to
deviations from the simple scaling (\ref{corral}). Nevertheless, we
show that the inter-event time distribution can be expressed in a universal
scaling form in terms of these four characteristic times. 
This scaling form  allows to better
enlighten the mechanism leading to universality for $D(\Delta t)$ 
and the deviations from it. The above analysis also
clarifies the dependence of $c$ on the mainshock magnitude for
intermediate mainshock sizes.


We assume that seismic occurrence can be modeled by a
time-dependent Poisson process with instantaneous rate
$\lambda(t)$. In this case, the inter-event time distribution for the
temporal interval $[0,T]$ is \cite{shc}

\begin{eqnarray}
D(\Delta t)&= &\frac{1}{N} \left [ \int_0^{T-\Delta t}
ds \lambda(s)\lambda(s+\Delta t)
\exp^{-\int _s^{s+\Delta t} du \lambda(u)} \right .
\nonumber
\\
&+&  \left .\lambda (\Delta t)
\exp^{-\int _0^{\Delta t} du \lambda(u)}\right ]
 \label{poi}
\end{eqnarray}

where $N=\int_0^T ds \lambda(s)$ is the number of events.
The widely accepted scenario is that seismic occurrence can be considered as
the superposition of Poissonian events occurring at constant rate $\mu$ and
independent aftershock sequences, which gives

\begin{equation}
\lambda(t)=\mu + \sum_{i:t_i<t} a_i 
(p-1) \left (\frac{t-t_i}{c_i}+1 \right)^{-p}
\label{lambda}
\end{equation}
where  $p > 1$ is the exponent of the
Omori law.  The quantity  $a_i$ is proportional to the rate 
of aftershocks correlated to  the $i$-th mainshock since, from
Eq.(\ref{lambda}), the total number of events triggered by the $i$-th
mainshock is $a_i c_i$. The productivity law \cite{hel}
indicates that $a_i$ is exponentially related to the mainshock
magnitude $a_i = A 10^{\alpha m_i}$. 
The  ETAS
model assumes $c_i=c$
whereas recent studies on experimental catalogs have obtained 
$c_i=c 10^{\alpha m_i}$ which leads to
the so-called generalized Omori law \cite{shc2,shc,lip1}. 
The dependence of $c_i$ on mainshock
magnitudes has been attributed to a dynamical scaling relation
involving time, space and energy \cite{lip,lip1,lip2,lip3} or  
to catalog incompleteness \cite{Kagan}. 
Inserting Eq.(\ref{lambda}) in Eq.(\ref{poi}), we obtain a
scaling form for $D(\Delta t)$ expressing time in unit of $1/\mu$,
\begin{equation}
D(\Delta t) =\mu G\left(\mu \Delta t,\mu/\overline a, \mu \overline c, \mu
T \right)
\label{scalD}
\end{equation}
where $\overline a$ ($\overline c$) is the value of $a_i$ ($c_i$)
averaged over all mainshocks. 
We show that Eq.(\ref{corral}) represents
a particular case of the more general scaling form (\ref{scalD}).
 Indeed, by definition, $R$ in the time interval $[0,T]$ is the inverse of
the average $\Delta t$, and from Eq.(\ref{scalD}) 
$R=\mu/ H\left(\mu/\overline a,\mu \overline c,\mu T\right) $,
with $H(y,z,w)=\int dx G(x,y,z,w)x$. Therefore, expressing $\mu$ in
terms of $R$ and setting ${\cal K}={\overline a}$ ${\overline c}$, we obtain 
\begin{equation}
D(\Delta t) =R G_1\left(R \Delta t,{\cal K}, \mu \overline c, \mu
T\right) .
\label{scalDD}
\end{equation}
For the ETAS model ${\cal K}=\overline a c$
is the branching ratio, i.e. the 
number of direct aftershocks per earthquake.

The complex form of Eq. (\ref{lambda}) does not allow
the full derivation of an analytical expression for $D(\Delta
t)$, unless one uses specific assumptions. SS
\cite{ss2}, for
instance, have exactly calculated $D(\Delta t)$ for $T \to \infty$ in the hypothesis that
each earthquake triggers, on average,  the same number of aftershocks,
i.e. $a_i=A$ and $c_i=c$. This solution exactly follows the scaling form
Eq.(\ref{scalDD}) with $G_1(x,y,z,\infty) =
  \exp\left (-x (1-y)-
\left (\omega^{2-p}-1\right)
z y /\big((1-y) (2-p)\big)\right)  \\
\left(\left(1-y+y\omega^{1-p}\right)^2
+(p-1)\omega^{-p} y (1-y)/z \right)$
with $\omega=1+(1-y)x/z$.
For a given choice of the parameters ${\cal K}$, $\mu c$ and $p$ the
above expression leads to a $D(\Delta t)$ in good agreement with the
experimental distribution. Interestingly, the above expression
coincides 
with the linear order
of the ETAS model expansion, in the limit $\Delta t \gg c$.
 Higher order terms lead to small differences with
the above solution \cite{ss3}.   

\begin{figure}
\includegraphics[width=8cm]{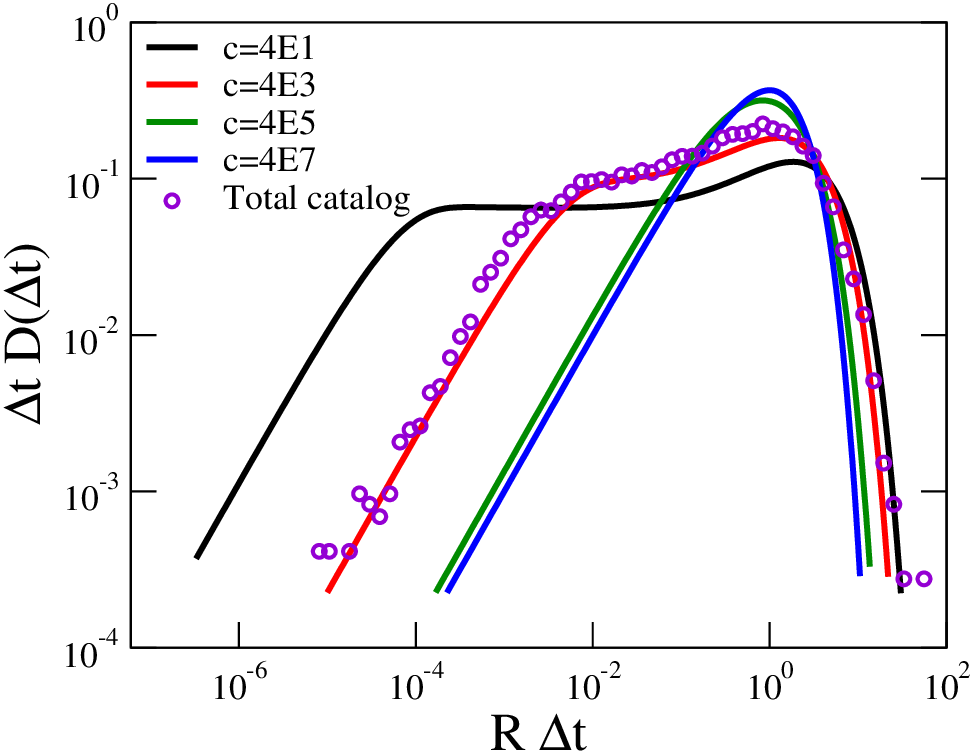}
\caption{\label{fig:figth}$\Delta t D(\Delta t)$ vs. $R \Delta t$
for the single Omori sequence 
with $\mu = 3 \cdot 10^{-7}$ events per
second and varying $c$ from $40$ to $4\cdot 10^7$ sec (from left to right). 
Other model parameters are $T=1 \cdot 10^7$ sec,
$A=1$ and $p=1.05$. Open circles represent $\Delta t D(\Delta t)$ for
the California catalog \cite{catalogo}. }
\end{figure}

A useful example to understand the role of the different time
scales in $D(\Delta t)$ can be obtained if we limit the calculation to 
events in a single aftershock sequence. 
A scaling form consistent with Eq.(\ref{scalDD}) has been already
obtained in ref.\cite{shc} assuming $\mu=0$. 
Here, we restrict to large $T$ and $\Delta t \ll T$  
assuming that $\lambda(t)$ is about constant in
$\Delta t$. 
This choice does not represent a loss of generality for sufficiently small
$\Delta t$. 
Under these assumptions, Eq.(\ref{poi})
becomes $D(\Delta t)= \frac{1}{N}  \int_0^T
ds \lambda^2(s)
\exp^{- \lambda(s) \Delta t}$
%
which can be analytically integrated.
 Three typical regimes can be identified: 
i) At large
times $\mu \Delta t > 1$ (i.e. $x>1$),with $\Delta t/T \ll1$, 
$D(\Delta t)$ decays as $\exp{(-\mu \Delta t)}$ as already observed \cite{mol}. 
ii) At
intermediate times $c < \Delta t <  1/\mu$ $(y<x<1)$, we observe a
power law decay $\Delta t^{-\epsilon}$  with $\epsilon=2-1/p$, as
predicted by Utsu \cite{uts}.  
iii) At small times $\Delta t \ll c$
$(x/y \ll 1)$, $D(\Delta t)$ becomes $\Delta t$
 independent. 
The three regimes can be
identified in Fig.~\ref{fig:figth} where,
 for a fixed value of $\mu$ and different $c$, we plot $\Delta t
D(\Delta t)$ vs $R \Delta t$. 
 This is equivalent to the representation adopted
by \cite{tou} and allows to better enlighten deviations from the scaling
relation (\ref{corral}). We observe that all curves present a
peak at $R \Delta t \simeq 1$ and then  exponentially decay for
$R\Delta t >1$. Conversely, at small times ($\Delta t<c$),
all the curves increase linearly since $D(\Delta t)$ is constant. 
The intermediate regime can be observed only for the two smallest
values of $c$, since only in these cases $\mu c \ll 1$ and the
 intermediate regime has a finite extension. In this regime
an about flat behavior is observed since for $p=1.05$, $\epsilon
\simeq 1$. 
Notice that one of the theoretical curves provides a
good qualitative fit of the distribution obtained from experimental
 data \cite{catalogo}.

\begin{figure}
\includegraphics[width=7cm]{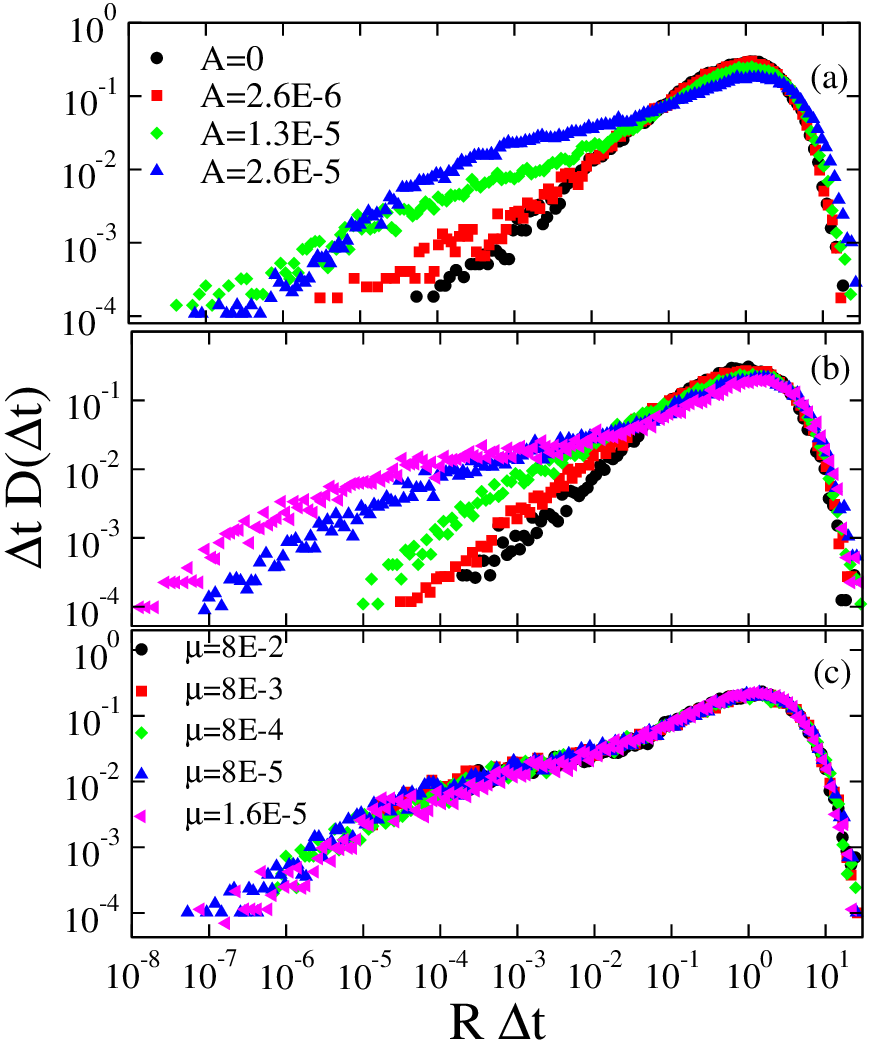}
\caption{\label{fig:intert1}$\Delta t D(\Delta t)$ vs. $R \Delta t$
  for the simulated catalogs with  $T=10^7$  and $p=1.2$.
 (a) $\Delta t D(\Delta t)$  for $c=3000$, $\mu=8 \cdot 10^{-6}$
events per unit time and varying ${\cal K}$ using different $A$ values
  reported in the legend.
(b) $\Delta t D(\Delta t)$  for $c=3000$,
${\cal K}=0.5$ and varying $\mu$ according to the legend in panel (c). (c)
We fix $c \mu=0.24$ obtaining data collapse with ${\cal K}=0.5$.
In all figures the time unit is the iteration step of the model.}
\end{figure}

In Fig.2 we present the results of numerical simulations of the 
ETAS model obtained
following the method of ref.\cite{hol}. 
We perform extensive simulations in
order to recover the limit $T\to\infty$ and neglect the dependence on
$\mu T$.
Previous studies \cite{hai} have proposed that $D(\Delta
t)$ only depends on the branching ratio and used this result to
measure the ratio 
between triggered and independent events in experimental
catalogs. Sornette et al. \cite{ss3} have shown 
that  the functional form of $D(\Delta t)$ also depends
on $p$, whose experimental values fluctuate around one.   
 Our numerical simulations with a random $p\in]1,1.6]$ indicate that
$D(\Delta t)$ depends on the average $p$ value. In the following we
present result for simulations with $p=1.2$. A very similar pattern is
obtained for other $p$ values.
Fig.~\ref{fig:intert1}a shows that for
fixed values of $\mu$ and $c$ the curves exhibit different behaviours
for different ${\cal K}$, in agreement with previous results.  
We then focus on the role of the parameter $c$.
In Fig.~\ref{fig:intert1}b, we plot $\Delta t D(\Delta t)$ at 
constant ${\cal K}$ and $c$ and for different values of $\mu$,
as in ref.\cite{tou}. We
confirm the existence of deviations from scaling (\ref{corral}) at
small $\Delta t$ which can be attributed to $c$. In order to
show that the dependence on $c$ enters in the scaling form as $\mu
\overline c$,
in Fig.~\ref{fig:intert1}c, we fix the branching ratio ${\cal K}$ and vary
$\mu$ as in Fig.~\ref{fig:intert1}b, but now $c$ is allowed to
vary keeping constant  $\mu c$. In this case all
the distributions collapse on the same master curve revealing an
"universal" behaviour also at small $\Delta t$. This result 
indicates that deviations from Eq.(\ref{corral}) rely on the presence
of the variable $\mu \overline c$ in Eq.(\ref{scalD}). The dependence on this
variable is relevant for $\Delta t<c$ and becomes negligible at larger
$\Delta t$.  This accounts for the appearance of deviations from
universality only at small $\Delta t$.

\begin{figure}
\includegraphics[width=9.cm]{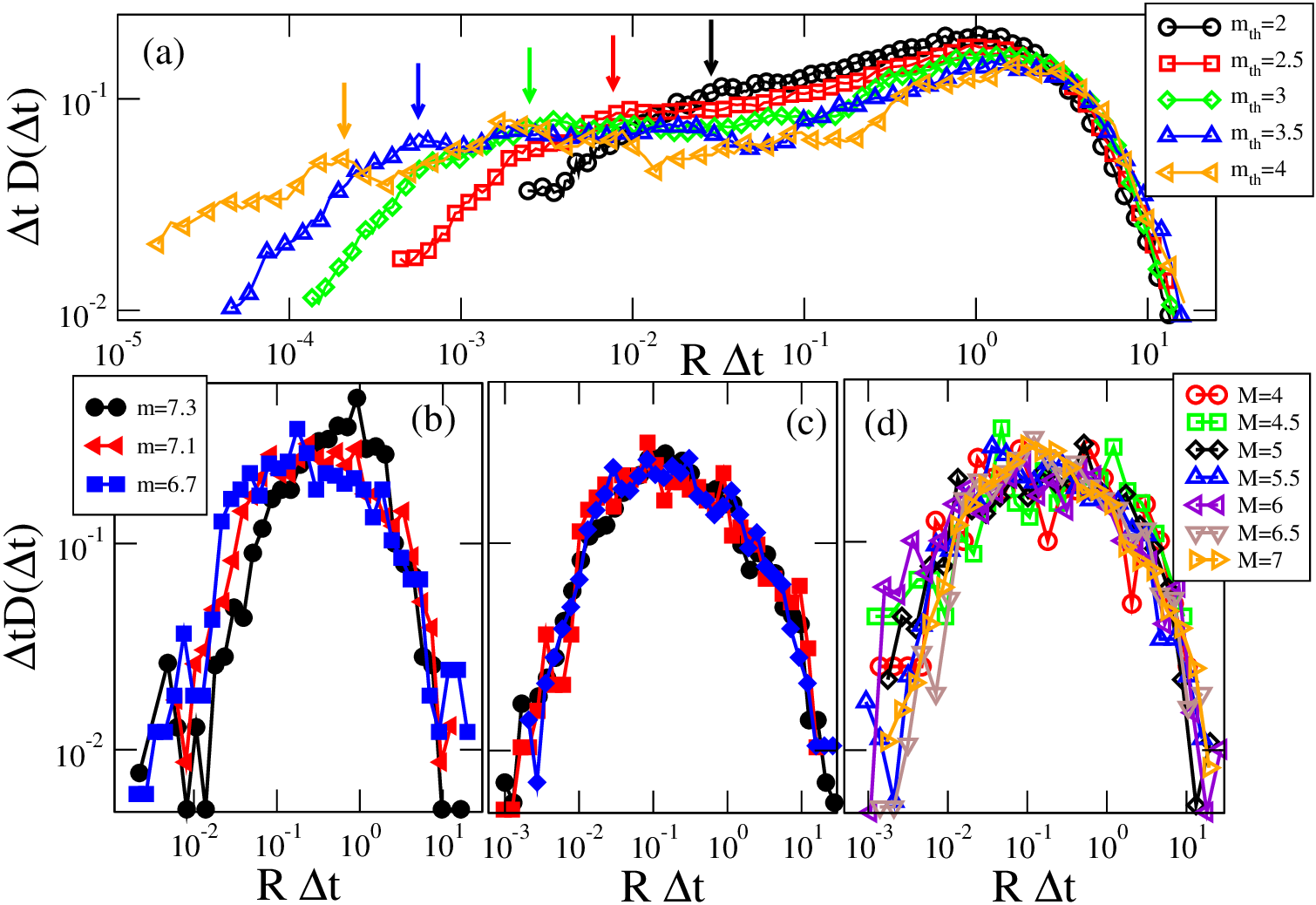}
\caption{\label{fig:fig2} (a) $\Delta t D(\Delta t)$ vs. $R \Delta t$ for
  different magnitude thresholds $m_{th}$; (b) for the 
Landers, Northridge and
Hector Mine sequences (mainshock magnitudes $m=7.3,6.7,7.1$, respectively)  
with fixed  duration $T=10$ days;
(c) for the 
Landers, Northridge and
Hector Mine sequences  
with $T$
chosen following the criterion of ref.\cite{bot};
(d) for all sequences with a mainshock magnitude
$m$ greater than 4 grouped in classes of $m$, $m \in [M,M+0.5[$.}
\end{figure}

We now explore the role of the characteristic time $c$ on
the scaling properties of $\Delta t D(\Delta t)$ for 
the experimental catalog \cite{catalogo}.
We first consider the whole catalog. In this case $\mu T$ is
very large and does not affect the scaling form (\ref{scalDD}). To isolate the
dependence on $\mu \overline{c}$, we plot  $\Delta t D(\Delta t)$
including in the analysis only earthquakes above a
lower magnitude thresholds $m_{th}$. $\mu/\overline a$, indeed, should
not depend on $m_{th}$ \cite{ss2} and $\Delta t D(\Delta t)$ is
expected to depend only on $R\Delta t$ and $\mu \overline{c}$. Fig.3a
shows deviations from universality.
 Since these deviations are
confined at small $\Delta t$ they are not easy to detect in the usual
plot $R^{-1} D(\Delta t)$ vs $R\Delta t$ \cite{cor}. Deviations must
be attributed to $\mu \overline{c}$ and allow to identify
$\overline c$ from the crossover points separating the linear growth
from the plateau (identified by arrows in Fig.3a). Using the   
known values of $R$ we obtain that $\overline{c}$ is  quite independent of
$m_{th}$. This is consistent with $c_i$ independent of mainshock
magnitudes but also with $c_i \propto 10^{\alpha (m_i-m_{th})}$ and
$\alpha<b$ \cite{nota}. The dependence of $c_i$ on $m_i$ and its
influence on the $D(\Delta t)$ can be obtained         
by restricting the
analysis to temporal periods soon after mainshocks. In these intervals,
$\mu$ can be neglected simplifying the scaling relation
(\ref{scalDD}). We further observe that for a single Omori sequence
$R=N/T=a_i c_i \left[1-(T/c_i)^{(1-p)}\right]/[T (p-1)]$ and therefore $R/a_i$
is a function only of the ratio $T/c_i$. In this case the scaling
simplifies  to  
\begin{equation}
D(\Delta t) =R G_2\left(R \Delta t, \frac{T}{c_i}\right ) .
\label{scalD3}
\end{equation}
We start by considering the main-aftershock sequences for the
three largest shocks recorded in the catalog:
 Landers, Northridge and Hector Mine. 
We consider as aftershocks all events with $m \ge 2.5$ occurring in a temporal window
$[0,T]$  after the main event and within the aftershock zone, i.e. a
radius $L=0.01\cdot 10^{0.5m_i}$km from the mainshock. Different
definitions of the aftershock zone \cite{chris} lead to very similar
results.
 We first fix $T=10$
days for all sequences. 
In Fig.~\ref{fig:fig2}b the curves do not collapse but show a
progressive shift as the mainshock magnitude increases.
This effect can be attributed to dependence of $c_i$ on
$m_i$. Therefore, at fixed $T$ the variable $T/c_i$ assumes different
values for each sequence violating the collapse  Eq.(\ref{corral}). 
As an alternative approach we  use the criterion proposed in ref.
\cite{bot} to identify the
end of a sequence:
namely a sequence ends when the rate $\lambda(t)$ reaches the
average Poisson rate $\mu \simeq 2$ events/day. 
Fig.~\ref{fig:fig2}c
clearly indicates a very good data collapse for the different
sequences in good agreement with Eq.(\ref{corral}).
This can be understood from the scaling relation (\ref{scalD3})
where the variable $T/c_i$ assumes the same value for all sequences. Indeed,
according to the Omori law, for $t \gg c$ the occurrence rate can be
expressed as $\lambda(t) \sim 10^{\alpha m_i}t^{-p}$ and 
the condition $\lambda(T)=2$ provides $T\sim 10^{\alpha m_i/p}$.
For the largest mainshocks, $c_i \sim 10^{b m_i/p}$ \cite{shc,bot} and  
therefore, the
condition $\alpha \simeq b/p$ implies that
the ratio $T/c_i$ is almost constant for the three sequences.

Next, we extend the above analysis to all sequences with mainshock
magnitude $m>4$. To improve the statistics, we group mainshocks in
classes of magnitude  $m \in [M,M+0.5[$.
Mainshocks are identified
with the criterion suggested in ref.\cite{bot} and the
duration is again fixed by the condition $\mu(T)=2$. Other methods \cite{chris}
for aftershock identification  provide similar results.
Fig.~\ref{fig:fig2}d shows  data collapse for all $M$
values. Since the criterion $\mu(T)=2$
roughly implies $T \sim 10^{\alpha M}$, the collapse of Fig.~\ref{fig:fig2}d, suggests that
the dependence of $c_i$ on the mainshock magnitude as $c_i \sim 10^{b m_i/p}$
is valid also for intermediate mainshock magnitudes.

In conclusion, we  address recent criticisms to the universal
behaviour of the inter-event distribution.
We follow the approach of ref.\cite{tou} and show that  $D(\Delta t)$ 
does exhibit universal features on the whole temporal range if four
characteristic time scales are taken into account.
In particular, deviations at small $\Delta t$ can be attributed to $c$
scaling differently from $\mu$. Whereas, 
by keeping
constant $T/c$ for different sequences, 
the $D(\Delta t)$  collapse onto a unique master curve. 

%
%






\begin{thebibliography}{}
\bibitem{nota} $\overline{c}=\int_{m_{th}}^{\infty}c_0
  10^{\alpha(m_i-m_{th})}P(m_i)/\int_{m_{th}}^{\infty}P(m_i)$. Using
  $P(m_i)=10^{-b m_i}$ we obtain $\overline{c}=c_0/(1-\alpha/b)$.
\end{thebibliography}
\end{document}